\documentclass[copyright,creativecommons,UKenglish,final]{eptcs}
\providecommand{\event}{TERMGRAPH
  2013} 

\usepackage{babel}
\usepackage[utf8]{inputenc}
\usepackage{tikz}

\usepackage{stmaryrd}
\usepackage{amssymb}
\usepackage{amsmath}
\usepackage{paralist}
\usepackage{color}
\usepackage{ifthen}

\usetikzlibrary{positioning,backgrounds,calc,decorations,arrows}
\usetikzlibrary{%
  decorations.pathmorphing,%
  decorations.pathreplacing,%
  decorations.markings,%
  decorations.shapes,%
  decorations.footprints,%
  decorations.fractals,%
  decorations.text%
}%
\tikzset{%
  >=latex',%
  single step/.style={>=to,->},%
  etc/.style={edge from parent/.style={-,dotted,thick,draw}},%
  shorten/.style={shorten >=#1, shorten <=#1},%
  level distance=1cm,%
  sibling distance = 10mm,%
  edge from parent/.style={->,draw},%
  fun/.style={>=to,->},%
  node name/.style args={#1:#2}{label={[black!70]#1:\small$#2$}}%
  }%

\newcommand\wsuc\oh

\newcommand\lebot{\le_\bot}

\newcommand\lebots{\lebot^\textsf{S}}

\newcommand\glb{\sqcap}
\newcommand\Lub{\bigsqcup}
\newcommand\Glb{\bigsqcap}

\newcommand\dd{\mathbf{d}}

\newcommand\dds{\dd_{\truncs{}{}}}

\newcommand\truncs[2]{#1\mathclose{\dagger}#2}

\newcommand\truncl[2]{#1{\setminus}#2}

\newcommand\reals{{\mathbb R}}

\newcommand\realsnn{\reals^+_0}
\renewcommand\epsilon{\varepsilon}

\newcommand\oh\widehat
\newcommand\ol\overline
\newcommand\ot\widetilde

\newcommand\canon[1]{\calC(#1)}

\newcommand\unrav[1]{\calU\left(#1\right)}

\newcommand\isom{\cong}

\newcommand\calC{\mathcal{C}}

\newcommand\calG{\mathcal{G}}

\newcommand\calP{\mathcal{P}}

\newcommand\calR{\mathcal{R}}

\newcommand\calT{\mathcal{T}}
\newcommand\calU{\mathcal{U}}
\newcommand\calV{\mathcal{V}}

\newcommand\calPos{\calP}

\mathchardef\mhyphen="2D

\newcommand\fcolon{\colon\,}
\newcommand\funto{\rightarrow}
\newcommand\limto{\rightarrow}
\newcommand\len[1]{\left\lvert #1 \right\rvert}

\newcommand\seq[1]{\langle #1 \rangle}
\newcommand\emptyseq{\seq{}}
\newcommand\srank[1]{\symb{ar}(#1)}

\newcommand\nin{\not\in}

\newcommand\prs{p}
\newcommand\mrs{m}

\newcommand\homto{\rightarrow}

\newcommand{\setcom}[2]{\set{#1\left\vert\vphantom{#1}\,#2\right.}}
\newcommand{\set}[1]{\left\{#1\right\}}

\newcommand\nodePos[2]{\calPos_{#1}(#2)}

\newcommand\depth[2]{\symb{depth}_{#1}(#2)}

\newcommand\iptgraphs[1][\Sigma]{\calG^\infty(#1_\bot)}

\newcommand\itgraphs[1][\Sigma]{\calG^\infty(#1)}
\newcommand\ctgraphs[1][\Sigma]{\calG_\calC(#1)}
\newcommand\ictgraphs[1][\Sigma]{\calG^\infty_\calC(#1)}
\newcommand\pctgraphs[1][\Sigma]{\calG_\calC(#1_\bot)}
\newcommand\ipctgraphs[1][\Sigma]{\calG^\infty_\calC(#1_\bot)}

\newcommand\pterms[1][\Sigma]{\calT(#1_\bot)}
\newcommand\ipterms[1][\Sigma]{\calT^\infty(#1_\bot)}
\newcommand\terms[1][\Sigma]{\calT(#1)}
\newcommand\iterms[1][\Sigma]{\calT^\infty(#1)}

\newcommand\symb[1]{\mathsf{#1}}
\newcommand\glab{\symb{lab}}
\newcommand\gsuc{\symb{suc}}

\newcommand\substAtPos[3]{#1[#3]_{#2}}

\newcommand\lhs[1]{#1_{l}}
\newcommand\rhs[1]{#1_{r}}

\newcommand\subgraph[2]{#1|_{#2}}

\def\nothing{}
\let\oldTo\to

\newcommand\preright{\mapsto}
\newcommand\finleftright{\leftrightarrow}
\newcommand\finright{\oldTo}
\newcommand\finleft{\leftarrow}
\newcommand\weakright{\hookrightarrow}
\newcommand\weakleft{\hookleftarrow}
\newcommand\strongright{\twoheadrightarrow}
\newcommand\strongleft{\twoheadleftarrow}
\newcommand\mrsright{\mathrel{\twoheadrightarrow^{\hspace{-11pt}m}\hspace{3pt}}}
\newcommand\mrsleft{\mathrel{\twoheadleftarrow^{\hspace{-7pt}m}\hspace{0pt}}}
\newcommand\prsright{\mathrel{\twoheadrightarrow^{\hspace{-9pt}p}\hspace{4pt}}}
\newcommand\prsleft{\mathrel{\twoheadleftarrow^{\hspace{-5pt}p}\hspace{0pt}}}

\newcommand\mrswright{\mathrel{\hookrightarrow^{\hspace{-10pt}m}\hspace{2pt}}}
\newcommand\mrswleft{\mathrel{\hookleftarrow^{\hspace{-7pt}m}\hspace{0pt}}}
\newcommand\prswright{\mathrel{\hookrightarrow^{\hspace{-9pt}p}\hspace{4pt}}}
\newcommand\prswleft{\mathrel{\hookleftarrow^{\hspace{-5pt}p}\hspace{0pt}}}

\newcommand{\RewArr}[2] {
  \RewStmt{#1}{\nothing}{#2}
}

\newcommand{\RewStmt}[3] {
  \def\RewArrArr{#1}
  \def\RewArrRhs{#2}
  \def\RewArrIter{#3}
  \RewArrI
}

\makeatletter \newcommand{\RewArrI}[1][\nothing] { \def\RewArrCxt{#1}
  \ifthenelse{\equal{\RewArrArr}{\oldTo} \OR
    \equal{\RewArrArr}{\preright} \OR
    \equal{\RewArrArr}{\finright} \OR
    \equal{\RewArrArr}{\finleftright} \OR
    \equal{\RewArrArr}{\prsright} \OR
    \equal{\RewArrArr}{\mrsright} \OR
    \equal{\RewArrArr}{\prswright} \OR
    \equal{\RewArrArr}{\mrswright} \OR
    \equal{\RewArrArr}{\strongright} \OR
    \equal{\RewArrArr}{\weakright}} {
    \RewArrArr\ifthenelse{\equal{\RewArrIter}{\nothing}}{}{^{\RewArrIter}}\ifthenelse{\equal{\RewArrCxt}{\nothing}}{}{_{\RewArrCxt}}
  } { \ifthenelse{\equal{\RewArrArr}{\finleft} \OR
      \equal{\RewArrArr}{\prsleft} \OR
      \equal{\RewArrArr}{\mrsleft} \OR
      \equal{\RewArrArr}{\prswleft} \OR
      \equal{\RewArrArr}{\mrswleft} \OR
      \equal{\RewArrArr}{\strongleft} \OR
      \equal{\RewArrArr}{\weakleft}}{
      \RewArrArr\ifthenelse{\equal{\RewArrIter}{\nothing}}{}{^{\RewArrIter}}\ifthenelse{\equal{\RewArrCxt}{\nothing}}{}{_{\RewArrCxt}}
    } { \latex@error{Rewrite arrow not defined}\@ehc } } \RewArrRhs }
\makeatother

\renewcommand{\to}{\RewArr{\finright}{\nothing}}

\newcommand{\fto}{\RewArr{\finright}}

\newcommand{\preto}{\RewArr{\preright}{\nothing}}

\newcommand{\pato}{\RewArr{\prsright}{\nothing}}

\newcommand{\mato}{\RewArr{\mrsright}{\nothing}}

\newcommand{\wmato}{\RewArr{\mrswright}{\nothing}}

\newcommand{\wpato}{\RewArr{\prswright}{\nothing}}

\newcommand\cons{{\,::\,}}

\usepackage{subfig}

\usepackage[strings]{underscore}
\usepackage[shortlabels,inline]{enumitem}
\usepackage{amsthm}

\theoremstyle{definition}
\newtheorem{definition}{Definition}[section]
\newtheorem{example}{Example}[section]

\theoremstyle{plain}
\newtheorem{theorem}{Theorem}[section]

\newtheorem{proposition}{Proposition}[section]

\tikzset{%
  level distance=8mm,%
  sibling distance = 8mm%
  }%

\usepackage[obeyDraft,colorinlistoftodos]{todonotes}
\newcommand\continue[1][]{\todo{{\bf continue here}}}%

\title{Convergence in Infinitary Term Graph Rewriting Systems is
  Simple (Extended Abstract)\thanks{The full version of this paper
    will appear in Mathematical Structures in Computer
    Science~\cite{bahr13mscs}.}}%
\author{Patrick Bahr
  \institute{Department of Computer Science, University of Copenhagen\\
    Universitetsparken 5, 2100 Copenhagen, Denmark}
  \email{paba@diku.dk} }
\def\titlerunning{Convergence in Infinitary Term Graph Rewriting Systems is
  Simple (Extended Abstract)}
\def\authorrunning{P. Bahr}
\begin{document}
\maketitle

\begin{abstract}
  In this extended abstract, we present a simple approach to
  convergence on term graphs that allows us to unify term graph
  rewriting and infinitary term rewriting. This approach is based on a
  partial order and a metric on term graphs. These structures arise as
  straightforward generalisations of the corresponding structures used
  in infinitary term rewriting.  We compare our simple approach to a
  more complicated approach that we developed earlier and show that
  this new approach is superior in many ways. The only unfavourable
  property that we were able to identify, viz.\ failure of full
  correspondence between weak metric and partial order convergence, is
  rectified by adopting a strong convergence discipline.
\end{abstract}

\section{Introduction}
\label{sec:introduction}

In \emph{infinitary term rewriting}~\cite{kennaway03book} we study
infinite terms and infinite rewrite sequences. Typically, this
extension to infinite structures is formalised by an ultrametric on
terms, which yields infinite terms by metric completion and provides a
notion of convergence to give meaning to infinite rewrite
sequences. In this paper we extend infinitary term rewriting to term
graphs. In addition to the metric approach, we also consider the
partial order approach to infinitary term rewriting~\cite{bahr10rta2}
and generalise it to the setting of term graphs.

One of the motivations for studying infinitary term rewriting is its
relation to \emph{non-strict evaluation}, which is used in programming
languages such as Haskell~\cite{marlow10haskell}. Non-strict
evaluation defers the evaluation of an expression until it is
``needed'' and thereby allows us to deal with conceptually infinite
data structures and computations. For example, the function
\texttt{from} defined below constructs for each number $n$ the
infinite list of consecutive numbers starting from $n$:
\begin{verbatim}
from(n) = n :: from(s(n))
\end{verbatim}
This construction is only conceptual and only results in a terminating
computation if it is used in a context where only finitely many
elements of the list are ``needed''. Infinitary term rewriting
provides us with an explicit limit construction to witness the outcome
of an infinite computation as it is, for example, induced by
\texttt{from}.  After translating the above function definition to a
term rewrite rule $\mathit{from}(x) \to x \cons \mathit{from}(s(x))$, we may derive an
infinite rewrite sequence
\[
\mathit{from}(0) \to 0 \cons \mathit{from}(s(0)) \to 0 \cons s(0)
\cons \mathit{from}(s(s(0))) \to \dots
\]
which converges to the infinite term $0 \cons s(0) \cons s(s(0)) \cons
\dots$, which represents the infinite list of numbers $0, 1, 2, \dots$
-- as intuitively expected.

Non-strict evaluation is rarely found in isolation, though. Usually,
it is implemented as lazy evaluation~\cite{henderson76popl}, which
complements a non-strict evaluation strategy with \emph{sharing}. The
latter avoids duplication of subexpressions by using pointers instead
of copying. For example, the function \texttt{from} above duplicates
its argument \texttt{n} -- it occurs twice on the right-hand side of
the defining equation. A lazy evaluator simulates this duplication by
inserting two pointers pointing to the actual argument.


While infinitary term rewriting is used to model the non-strictness of
lazy evaluation, term graph rewriting models the sharing part of it.
By endowing term graph rewriting with a notion of convergence like in
infinitary term rewriting, we aim to unify the two formalisms into one
calculus, thus allowing us to model both aspects within the same
calculus.

\paragraph{Contributions \& Outline}
\label{sec:contributions}
At first we recall the basic notions of infinitary term rewriting
(Section~\ref{sec:infin-term-rewr}). Afterwards, we construct a metric
and a partial order on term graphs and show that both are suitable as
a basis for notions of convergence in term graph rewriting
(Section~\ref{sec:graphs-term-graphs}). Based on these structures we
introduce notions of convergence (weak and strong variants) for term
graph rewriting and show correspondences between metric-based and
partial order-based convergence (Section~\ref{sec:weak-convergence}
and \ref{sec:strong-convergence}). We then present soundness and
completeness properties of the resulting infinitary term graph
rewriting calculi w.r.t.\ infinitary term rewriting
(Section~\ref{sec:soundness}). Lastly, we compare our calculi with
previous approaches (Section~\ref{sec:concluding-remarks}).

\section{Infinitary Term Rewriting}
\label{sec:infin-term-rewr}

Before starting with the development of infinitary \emph{term graph}
rewriting, we recall the basic notions of infinitary \emph{term}
rewriting. Rewrite sequences in infinitary rewriting, also called
\emph{reductions}, are sequences of the form
$(\phi_\iota)_{\iota<\alpha}$, where each $\phi_\iota$ is a rewrite
step from a term $t_\iota$ to $t_{\iota+1}$ in a term rewriting system
(TRS) $\calR$, denoted $\phi_\iota\fcolon t_\iota \to[\calR]
t_{\iota+1}$. The length $\alpha$ of such a sequence can be an
arbitrary ordinal. For example, the infinite reduction indicated in
Section~\ref{sec:introduction} is the sequence
$(\phi^\mathrm{f}_i\fcolon t^\mathrm{f}_i \to[\calR^\mathrm{f}]
t^\mathrm{f}_{i+1})_{i<\omega}$, where $t^\mathrm{f}_i = 0 \cons \dots
\cons s^{i-1}(0) \cons \mathit{from}(s^i(0))$ for all $i<\omega$ and
$\calR^\mathrm{f}$ is the TRS consisting of the single rule $\mathit{from}(x)
\to x \cons \mathit{from}(s(x))$.

\subsection{Metric Convergence}
\label{sec:metric-convergence}

The above definition of reductions ensures that consecutive rewrite
steps are ``compatible'', i.e.\ the result term of the $\iota$-th
step, viz.\ $t_{\iota+1}$, is the start term of the $(\iota+1)$-st
step. However, this definition does not relate the start terms of
steps at limit ordinal positions to the terms that preceded it. For
example, we can extend the abovementioned reduction
$(\phi^\mathrm{f}_i)_{i<\omega}$ of length $\omega$, to a reduction
$(\phi^\mathrm{f}_i)_{i<\omega+1}$ of length $\omega +1$ using
any reduction step $\phi^\mathrm{f}_\omega$,
e.g. $\phi^\mathrm{f}_\omega\fcolon \mathit{from}(0) \to 0 \cons
\mathit{from}(s(0))$. In our informal notation this reduction
$(\phi^\mathrm{f}_i)_{i<\omega+1}$ reads as follows:
\[
\mathit{from}(0) \to 0 \cons \mathit{from}(s(0)) \to 0 \cons s(0) \cons \mathit{from}(s(s(0)))
\to \quad \dots\quad \mathit{from}(0) \to 0 \cons \mathit{from}(s(0))
\]
Intuitively, this does not make sense since the sequence of terms that
precedes the last step intuitively converge to the term $0 \cons s(0)
\cons s(s(0)) \cons \dots$, but not $\mathit{from}(0)$.

In infinitary term rewriting such reductions are ruled out by a notion
of convergence and a notion of continuity that follows from
it. Typically, this notion of convergence is derived from a metric
$\dd$ on the set of (finite and infinite) terms $\iterms$: $\dd(s,t) =
0$ if $s = t$, and $\dd(s,t) = 2^{-d}$ otherwise, where $d$ is the
minimal depth at which $s$ and $t$ differ. Using this metric, we may
also construct the set of (finite and infinite) terms $\iterms$ by
\emph{metric completion} of the metric space $(\terms,\dd)$ of finite
terms.

The mode of convergence in the metric space $(\iterms,\dd)$ is the
basis for the notion of \emph{weak $\mrs$-convergence} of reductions:
a reduction $S = (\phi_\iota\fcolon t_\iota \to[\calR]
t_{\iota+1})_{\iota<\alpha}$ is \emph{weakly $\mrs$-continuous} if
$\lim_{\iota\limto\lambda} t_\iota = t_\lambda$ for all limit ordinals
$\lambda < \alpha$; it \emph{weakly $\mrs$-converges} to a term $t$,
denoted $S\fcolon t_0 \wmato[\calR] t$, if it is weakly
$\mrs$-continuous and $\lim_{\iota\limto\wsuc\alpha} t_\iota = t$,
where $\wsuc\alpha$ is the length of the underlying sequence of terms
$(t_\iota)_{\iota<\wsuc\alpha}$. For example, the reduction
$(\phi^\mathrm{f}_i)_{i<\omega}$ weakly $\mrs$-converges to the term
$0 \cons s(0) \cons s(s(0)) \cons \dots$; but the sequence
$(\phi^\mathrm{f}_i)_{i<\omega+1}$ does not weakly $\mrs$-converge, it
is not even weakly $\mrs$-continuous as
$\lim_{\iota\limto\omega}t^\mathrm{f}_\iota$ is not $\mathit{from}(0)$.

Weak $\mrs$-convergence is quite a general notion of convergence. For
example, given a rewrite rule $a \to a$, we may derive the reduction
$a \to a \to \dots$, which weakly $\mrs$-converges to $a$ even though
the rule $a \to a$ is applied again and again at the same
position. This generality causes many desired properties to break,
such as unique normal form properties and
compression~\cite{kennaway95ic}. That is why Kennaway et
al.~\cite{kennaway95ic} introduced \emph{strong $\mrs$-convergence},
which in addition requires that the depth at which rewrite steps take
place tends to infinity as one approaches a limit ordinal: Let $S =
(\phi_\iota\fcolon t_\iota \to[\pi_\iota] t_{\iota+1})_{\iota<\alpha}$
be a reduction, where each $\pi_\iota$ indicates the position at which
the step $\phi_\iota$ takes place and $\len{\pi_\iota}$ denotes the
length of the position $\pi_\iota$. The reduction $S$ is said to be
\emph{strongly $\mrs$-continuous} (resp.\ \emph{strongly
  $\mrs$-converge} to $t$, denoted $S\fcolon t_0 \mato t$) if it is
weakly $\mrs$-continuous (resp.\ weakly $\mrs$-converges to $t$) and
if $(\len{\pi_\iota})_{\iota<\lambda}$ tends to infinity for all limit
ordinals $\lambda < \alpha$ (resp.\ $\lambda \le \alpha$).  For
example, the reduction $(\phi^\mathrm{f}_i)_{i<\omega}$ also strongly
$\mrs$-converges to the term $0 \cons s(0) \cons s(s(0)) \cons
\dots$. On the other hand, the reduction $a \to a \to \dots$ indicated
above weakly $\mrs$-converges to $a$, but it does not strongly
$\mrs$-converge to $a$.

\subsection{Partial Order Convergence}
\label{sec:part-order-conv}

Alternatively to the metric approach illustrated in
Section~\ref{sec:metric-convergence}, convergence can also be
formalised using a partial order $\lebot$ on terms. The idea to use
this partial order for infinitary rewriting goes back to
Corradini~\cite{corradini93tapsoft}. The signature $\Sigma$ is
extended to the signature $\Sigma_\bot$ by adding a fresh constant
symbol $\bot$. When dealing with terms in $\ipterms$, we call terms
that do not contain the symbol $\bot$, i.e.\ terms that are contained
in $\iterms$, \emph{total}. We define $s \lebot t$ iff $s$ can be
obtained from $t$ by replacing some subterm occurrences in $t$ by
$\bot$. Interpreting the term $\bot$ as denoting ``undefined'',
$\lebot$ can be read as ``is less defined than''. The pair
$(\ipterms,\lebot)$ is known to form a \emph{complete semilattice}
\cite{goguen77jacm}, i.e.\ it has a least element $\bot$, each
directed set $D$ in $(\ipterms,\lebot)$ has a \emph{least upper bound}
(\emph{lub}) $\Lub D$, and every \emph{non-empty} set $B$ in
$(\ipterms,\lebot)$ has \emph{greatest lower bound} (\emph{glb}) $\Glb
B$. In particular, this means that for any sequence
$(t_\iota)_{\iota<\alpha}$ in $(\ipterms,\lebot)$, its \emph{limit
  inferior}, defined by $\liminf_{\iota \limto \alpha}t_\iota =
\Lub_{\beta<\alpha} \left(\Glb_{\beta \le \iota < \alpha}
  t_\iota\right)$, exists.

In the same way that the limit in the metric space gives rise to weak
$\mrs$-continuity/-convergence, the limit inferior gives rise to
\emph{weak $\prs$-continuity} and \emph{weak $\prs$-convergence};
simply replace $\lim$ by $\liminf$. We write $S\fcolon t_0 \wpato t$
if a reduction $S$ starting with term $t_0$ weakly $\prs$-converges to
$t$. The defining difference between the two approaches is that
$\prs$-continuous reductions always $\prs$-converge. The reason for
that lies in the complete semilattice structure of
$(\ipterms,\lebot)$, which guarantees that the limit inferior always
exists (in contrast to the limit in a metric space).

The definition of the strong variant of $\prs$-convergence is a bit
different from the one of $\mrs$-convergence, but it follows the same
idea: a reduction $(\phi_i\fcolon t_i\to[\pi_i] t_{i+1})_{i<\omega}$
weakly $\mrs$-converges iff the minimal depth $d_i$ at which two
consecutive terms $t_i, t_{i+1}$ differ tends to infinity. The strong
variant of $\mrs$-convergence is a conservative approximation of this
condition; it requires $\len{\pi_i}$ to tend to infinity. This
approximation is conservative since $\len{\pi_i} \le d_i$; differences
between consecutive terms can only occur below the position at which a
rewrite rule was applied.

In the partial order approach we can make this approximation more
precise since we have the whole term structure at our disposal instead
of only the measure provided by the metric $\dd$. In the case of
$\mrs$-convergence, we replaced the actual depth of a minimal
difference $d_i$ with its conservative under-approximation
$\len{\pi_i}$. For $\prs$-convergence, we replace the glb $t_i \glb
t_{i+1}$, which intuitively represents the common information shared
by $t_i$ and $t_{i+1}$, with the conservative under-approximation
$\substAtPos{t_i}{\pi_i}{\bot}$, which replaces the redex at position
$\pi_i$ in $t_i$ with $\bot$. This term
$\substAtPos{t_i}{\pi_i}{\bot}$ -- called the \emph{reduction context}
of the step $\phi_i\fcolon t_i \to t_{i+1}$ -- is a lower bound of
$t_i$ and $t_{i+1}$ w.r.t.\ $\lebot$ and is, thus, also smaller than
$t_i \glb t_{i+1}$. The definition of strong $\prs$-convergence is
obtained from the definition of weak $\prs$-convergence by replacing
$\liminf_{\iota\limto\lambda} t_\iota$ with
$\liminf_{\iota\limto\lambda} \substAtPos{t_\iota}{\pi_\iota}{\bot}$.

A reduction $S = (\phi_\iota\fcolon t_\iota \to[\pi_\iota]
t_{\iota+1})_{\iota<\alpha}$ is called \emph{strongly
  $\prs$-continuous} if $\liminf_{\iota\limto\lambda}
\substAtPos{t_i}{\pi_i}{\bot} = t_\lambda$ for all limit ordinals
$\lambda < \alpha$; it \emph{strongly $\prs$-converges} to $t$,
denoted $S\fcolon t_0 \pato t$, if it is strongly $\prs$-continuous
and either $\liminf_{\iota\limto\alpha} \substAtPos{t_i}{\pi_i}{\bot}
= t$ in case $\alpha$ is a limit ordinal, or $t = t_{\alpha+1}$
otherwise.

\begin{example}
  The previously mentioned reduction $(\phi^\mathrm{f}_i)_{i<\omega}$
  both strongly and weakly $\prs$-converges to the infinite term $0
  \cons s(0) \cons s(s(0)) \cons \dots$ -- like in the metric
  approach. However, while the reduction $a \to a \to \dots$ does not
  strongly $\mrs$-converge, it strongly $\prs$-converges to the term
  $\bot$.
\end{example}

The partial order approach has some advantages over the metric
approach. As explained above, every $\prs$-continuous reduction is
also $\prs$-convergent. Moreover, strong $\prs$-convergence has some
properties such as infinitary normalisation and infinitary confluence
of orthogonal systems~\cite{bahr10rta2} that are not enjoyed by strong
$\mrs$-convergence.

Interestingly, however, the partial order-based notions of convergence
are merely conservative extensions of the metric-based ones:
\begin{theorem}[\cite{bahr09master,bahr10rta2}] 
  \label{thr:strongExt}
  For every reduction $S$ in a TRS, the following equivalences hold:
  \begin{center}
    \begin{enumerate*}[(i)]
    \item $S\fcolon s \wmato t$ \quad iff \quad
      $S\fcolon s \wpato t$ in $\iterms$.\label{item:strongExtI}%
      \hspace{8mm}
    \item $S\fcolon s \mato t$ \quad iff \quad
      $S\fcolon s \pato t$ in $\iterms$.\label{item:strongExtII}%
    \end{enumerate*}
  \end{center}
\end{theorem}
The phrase ``in $\iterms$'' means that all terms in $S$ are total
(including $t$). That is, if restricted to total terms, $\mrs$- and
$\prs$-convergence coincide.

\section{Graphs and Term Graphs}
\label{sec:graphs-term-graphs}

In this section, we present our notion of term graphs and generalise
the metric $\dd$ and the partial order $\lebot$ from terms to term
graphs.

Our notion of graphs and term graphs is largely taken from Barendregt et
al.~\cite{barendregt87parle}.
\begin{definition}[graphs]
  \label{def:graph}
  A \emph{graph} over signature $\Sigma$ is a triple $g =
  (N,\glab,\gsuc)$ consisting of a set $N$ (of \emph{nodes}), a
  \emph{labelling function} $\glab\fcolon N \funto \Sigma$, and a
  \emph{successor function} $\gsuc\fcolon N \funto N^*$ such that
  $\len{\gsuc(n)} = \srank{\glab(n)}$ for each node $n\in N$, i.e.\ a
  node labelled with a $k$-ary symbol has precisely $k$ successors. If
  $\gsuc(n) = \seq{n_0,\dots,n_{k-1}}$, then we write $\gsuc_{i}(n)$
  for $n_i$.
\end{definition}
The successor function $\gsuc$ defines, for each node $n$, directed
edges from $n$ to $\gsuc_i(n)$.  A path from a node $m$ to a node $n$
is a finite sequence $\seq{e_0,\dots,e_l}$ of numbers such that
$n=\gsuc_{e_l}(\dots \gsuc_{e_0}(m))$, i.e.\ $n$ is reached from $m$
by taking the $e_0$-th edge, then the $e_1$-th edge etc.

\begin{definition}[term graphs]
  \label{def:tgraph}
  A \emph{term graph} $g$ over $\Sigma$ is a tuple $(N,\glab,\gsuc,r)$
  consisting of an \emph{underlying} graph $(N,\glab,\gsuc)$ over
  $\Sigma$ whose nodes are all reachable from the \emph{root node}
  $r\in N$. The class of all term graphs over $\Sigma$ is denoted
  $\itgraphs$. A \emph{position} of $n \in N$ in $g$ is a path in the
  underlying graph of $g$ from $r$ to $n$. The set of all positions of
  $n$ in $g$ is denoted $\nodePos{g}{n}$. The \emph{depth} of $n$ in
  $g$, denoted $\depth{g}{n}$, is the minimum of the lengths of the
  positions of $n$ in $g$, i.e.\ $\depth{g}{n} = \min
  \setcom{\len{\pi}}{\pi \in \nodePos{g}{n}}$. The term graph $g$ is
  called a \emph{term tree} if each node in $g$ has exactly one
  position. We use the notation $N^{g}$, $\glab^{g}$, $\gsuc^{g}$ and
  $r^{g}$ to refer to the respective components $N$,$\glab$, $\gsuc$
  and $r$ of $g$. Given a graph or a term graph $h$ and a node $n$ in
  $h$, we write $\subgraph{h}{n}$ to denote the sub-term graph of $h$
  rooted in $n$.
\end{definition}

The notion of homomorphisms is crucial for dealing with term
graphs. For greater flexibility, we will parametrise this notion by a
set of constant symbols $\Delta$ for which the homomorphism condition
is suspended. This will allow us to deal with variables and partiality
appropriately.
\begin{definition}[$\Delta$-homomorphisms]
  \label{def:D-hom}
  Let $\Sigma$ be a signature, $\Delta\subseteq \Sigma^{(0)}$, and
  $g,h \in \itgraphs$. A \emph{$\Delta$-homomorphism} $\phi$ from $g$
  to $h$, denoted $\phi\fcolon g \homto_\Delta h$, is a function
  $\phi\fcolon N^g \funto N^h$ with $\phi(r^g) = r^h$ that satisfies
  the following equations for all for all $n \in N^g$ with $\glab^g(n)
  \nin \Delta$:
  \begin{align*}
    \glab^g(n) &= \glab^h(\phi(n))
    \tag{labelling}\\
    \phi(\gsuc^g_i(n)) &= \gsuc^h_i(\phi(n)) \quad \text{ for all } 0 \le i <
    \srank{\glab^g(n)} \tag{successor}
  \end{align*}
\end{definition}

Note that, for $\Delta = \emptyset$, we get the usual notion of
homomorphisms on term graphs (e.g.\ Barendsen~\cite{barendsen03book})
and from that the notion of isomorphisms. The nodes labelled with
symbols in $\Delta$ can be thought of as holes in the term graphs that
can be filled with other term graphs.

We do not want to distinguish between isomorphic term
graphs. Therefore, we use a well-known trick~\cite{plump99hggcbgt} to
obtain canonical representatives of isomorphism classes of term
graphs.
\begin{definition}
  \label{def:canTgraph}
  A term graph $g$ is called \emph{canonical} if $n = \nodePos{g}{n}$
  holds for each $n \in N^g$. That is, each node is the set of its
  positions in the term graph. The set of all (finite) canonical term
  graphs over $\Sigma$ is denoted $\ictgraphs$ (resp.\
  $\ctgraphs$). For each term graph $h \in \ictgraphs$, its
  \emph{canonical representative} $\canon{h}$ is obtained from $h$ by
  replacing each node $n$ in $h$ by $\nodePos{h}{n}$.

\end{definition}

This construction indeed yields a canonical representation of
isomorphism classes. More precisely: $g \isom \canon g$ for all
$g\in\itgraphs$, and $g \isom h$ iff $\canon g = \canon h$ for all
$g,h \in \itgraphs$.

We consider the set of terms $\iterms$ as the subset of canonical term
trees of $\ictgraphs$.  With this correspondence in mind, we can
define the \emph{unravelling} of a term graph $g$ as the unique term
$\unrav g$ such that there is a homomorphism $\phi\fcolon \unrav g
\homto g$. For example, $g_0$ from Figure~\ref{fig:convWeird} is the
unravelling of $g_1$, and $h_0$ and $g_\omega$ from
Figure~\ref{fig:fixedPointComb} both unravel to the infinite term
$@(f,@(f,\dots))$. Term graphs that unravel to the same term are
called \emph{bisimilar}.


\subsection{A Simple Partial Order on Term Graphs}
\label{sec:simple-partial-order}

In this section, we want to establish a partial order suitable for
formalising convergence of sequences of canonical term graphs
similarly to weak $\prs$-convergence on terms.

Weak $\prs$-convergence on term rewriting systems is based on the
partial order $\lebot$ on $\ipterms$, which instantiates occurrences
of $\bot$ from left to right, i.e.\ $s \lebot t$ iff $t$ is obtained
by replacing occurrences of $\bot$ in $s$ by arbitrary terms in
$\ipterms$.  Analogously, we consider the class of \emph{partial term
  graphs} simply as term graphs over the signature $\Sigma_\bot =
\Sigma \uplus \set{\bot}$. In order to generalise the partial order
$\lebot$ to term graphs, we need to formalise the instantiation of
occurrences of $\bot$ in term graphs. For this purpose, we shall use
$\Delta$-homomorphisms with $\Delta=\set\bot$, or $\bot$-homomorphisms
for short. A $\bot$-homomorphism $\phi\colon g \to_\bot h$ maps each
node in $g$ to a node in $h$ while ``preserving its
structure''. Except for nodes labelled $\bot$ this also includes
preserving the labelling. This exception to the homomorphism condition
allows the $\bot$-homomorphism $\phi$ to instantiate each $\bot$-node
in $g$ with an arbitrary node in $h$.  Using $\bot$-homomorphisms, we
arrive at the following definition for our simple partial order
$\lebots$ on term graphs:
\begin{definition}
  For each $g,h \in \ipctgraphs$, define $g \lebots h$ iff there is
  some $\phi\fcolon g \homto_\bot h$.
\end{definition}

One can verify that $\lebots$ indeed generalises the partial order
$\lebot$ on terms. Considering terms as canonical term trees, we
obtain the following characterisation of $\lebot$ on terms $s,t\in
\ipterms$:
\[
s \lebot t \iff \text{ there is a $\bot$-homomorphism } \phi\fcolon s
\homto_\bot t.
\]

The first important result for $\lebots$ is that the semilattice
structure that we already had for $\lebot$ is preserved by this
generalisation:
\begin{theorem}
  \label{thr:complSemilattice}
  The partially ordered set $(\ipctgraphs,\lebots)$ forms a complete
  semilattice.
\end{theorem}

For terms, we already know that the set of (potentially infinite)
terms can be constructed by forming the \emph{ideal completion} of the
partially ordered set $(\pterms,\lebot)$ of finite terms
\cite{berry77popl}. More precisely, the ideal completion of
$(\pterms,\lebot)$ is order isomorphic to $(\ipterms,\lebot)$.

An analogous result can be shown for term graphs:
\begin{theorem}
  \label{thr:idealCompletion}
  The ideal completion of $(\pctgraphs, \lebots)$ is order isomorphic
  to $(\ipctgraphs,\lebots)$.
\end{theorem}

\subsection{A Simple Metric on Term Graphs}
\label{sec:simple-metric-term}

Next, we shall generalise the metric $\dd$ from terms to term
graphs. To achieve this, we need to formalise what it means for two
term graphs to coincide up to a certain depth, so that we can
reformulate the definition of the metric $\dd$ for term graphs. To
this end, we follow the same idea that the original definition of
$\dd$ on terms from Arnold and Nivat~\cite{arnold80fi} was based
on. In particular, we introduce a truncation construction that cuts
off nodes below a certain depth:
\begin{definition}
  \label{def:trunca}
  Let $g \in \iptgraphs$ and $d \le \omega$. The \emph{simple
    truncation} $\truncs{g}{d}$ of $g$ at $d$ is the term graph
  defined as follows:
  \begin{align*}
    N^{\truncs{g}{d}} &= \setcom{n \in N^g}{\depth{g}{n} \le d}
    & r^{\truncs{g}{d}} &= r^g
    \\
    \glab^{\truncs{g}{d}}(n) &= 
    \begin{cases}
      \glab^g(n) &\text{if } \depth{g}{n} < d \\
      \bot  &\text{if } \depth{g}{n} = d
    \end{cases} &
    \gsuc^{\truncs{g}{d}}(n) &=
    \begin{cases}
      \gsuc^g(n) &\text{ if }\depth{g}{n} < d\\
      \emptyseq &\text{ if }\depth{g}{n} = d
    \end{cases}
  \end{align*}
\end{definition}

The definition of the simple metric $\dds$ follows straightforwardly:
\begin{definition}
  The \emph{simple distance} $\dds\fcolon \ictgraphs \times \ictgraphs
  \to \realsnn$ is defined as follows:
  \begin{gather*}
    \dds(g,h) =
    \begin{cases}
      0&\text{if } g = h\\%
      2^{-d}&\text{if } g \neq h \text{ and } d
      =\max\setcom{e<\omega}{\truncs{g}{e}\isom\truncs{h}{e}}
    \end{cases}
  \end{gather*}
\end{definition}

Again, we can verify that $\dds$ generalises $\dd$. In particular, we
can show that our notion of truncation coincides with that of Arnold
and Nivat~\cite{arnold80fi} if restricted to terms.

As desired, this generalisation retains the complete ultrametric space
structure:
\begin{theorem}
  \label{thr:smetricComplete}%
  The pair $(\ictgraphs,\dds)$ forms a complete ultrametric space.
\end{theorem}

The metric space analogue to ideal completion is metric completion. On
terms, we already know that we can construct the set of (potentially
infinite) terms $\iterms$ by metric completion of the metric space
$(\terms,\dd)$ of finite terms \cite{barr93tcs}. More precisely, the
metric completion of $(\terms,\dd)$ is the metric space
$(\iterms,\dd)$. This property generalises to term graphs as well:
\begin{theorem}
  \label{thr:metricCompletion}
  The metric completion of $(\ctgraphs,\dds)$ is the metric space
  $(\ictgraphs,\dds)$.
\end{theorem}

\begin{figure}
  \centering
  \begin{tikzpicture}[node distance=15mm]%
    \node (r1) {$f$} %
    child{ node (n1) {$c$} }%
    child{ node (n2) {$c$} };%
    \node[right=of r1] (r2) {$f$}%
    child {%
      node (n2) {$c$}%
      edge from parent[transparent] %
    };%
    \draw[->] (r2)%
    edge [bend right=25] (n2)%
    edge [bend left=25] (n2);%
    
    \draw[single step,shorten=5mm] (r1) -- (r2);%
    
    \node[right=of r2] (r3) {$f$} %
    child{ node (n1) {$c$} }%
    child{ node (n2) {$c$} };%
    
    \draw[single step,shorten=5mm] (r2) -- (r3);%
    
    \node[right=of r3] (r4) {$f$}%
    child {%
      node (n2) {$c$}%
      edge from parent[transparent] %
    };%
    \draw[->] (r4)%
    edge [bend right=25] (n2)%
    edge [bend left=25] (n2);%
    \draw[single step,shorten=5mm] (r3) -- (r4);%

    \node[node distance=25mm,right=of r4] (r5) {$f$} %
    child{ node (n1) {$c$} }%
    child{ node (n2) {$c$} };%
    
    \draw[dotted,thick,shorten=10mm] (r4) -- (r5);%
    \begin{scope}[node distance=7mm]
      \node[below=of r1] {$(g_0)$};
      \node[below=of r2] {$(g_1)$};
      \node[below=of r3] {$(g_2)$};
      \node[below=of r4] {$(g_4)$};
      \node[below=of r5] {$(g_\omega)$};
    \end{scope}
  \end{tikzpicture}
  \caption{Limit inferior in the presence of acyclic sharing.}
  \label{fig:convWeird}
\end{figure}
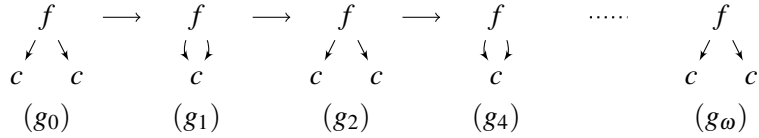

\section{Infinitary Term Graph Rewriting}
\label{sec:infin-term-graph}

In this paper, we adopt the term graph rewriting framework of
Barendregt et al.~\cite{barendregt87parle}. In order to represent
placeholders in rewrite rules, we use variables -- in a manner much
similar to term rewrite rules. To this end, we consider a signature
$\Sigma_\calV = \Sigma\uplus\calV$ that extends the signature $\Sigma$
with a set $\calV$ of nullary variable symbols.
\begin{definition}[term graph rewriting systems]
  Given a signature $\Sigma$, a \emph{term graph rule} $\rho$ over
  $\Sigma$ is a triple $(g,l,r)$ where $g$ is a graph over
  $\Sigma_\calV$ and $l,r \in N^g$ such that all nodes in $g$ are
  reachable from $l$ or $r$. We write $\lhs\rho$ resp.\ $\rhs\rho$ to
  denote the left- resp.\ right-hand side of $\rho$, i.e.\ the term
  graph $\subgraph{g}{l}$ resp.\ $\subgraph{g}{r}$. Additionally, we
  require that for each variable $v\in\calV$ there is at most one node
  $n$ in $g$ labelled $v$, and we have that $n \neq l$ and that $n$ is
  reachable from $l$ in $g$.  A \emph{term graph rewriting system
    (GRS)} $\calR$ is a pair $(\Sigma,R)$ with $\Sigma$ a signature
  and $R$ a set of term graph rules over $\Sigma$.
\end{definition}

The notion of unravelling straightforwardly extends to term graph
rules: the \emph{unravelling} of a term graph rule $\rho$, denoted
$\unrav{\rho}$, is the term rule $\unrav{\rho_l} \to
\unrav{\rho_r}$. The unravelling of a GRS $\calR=(\Sigma,R)$, denoted
$\unrav{\calR}$, is the TRS $(\Sigma,\setcom{\unrav{\rho}}{\rho\in
  R})$.

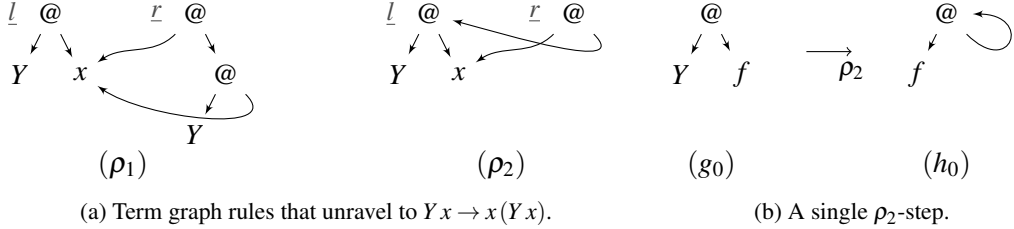
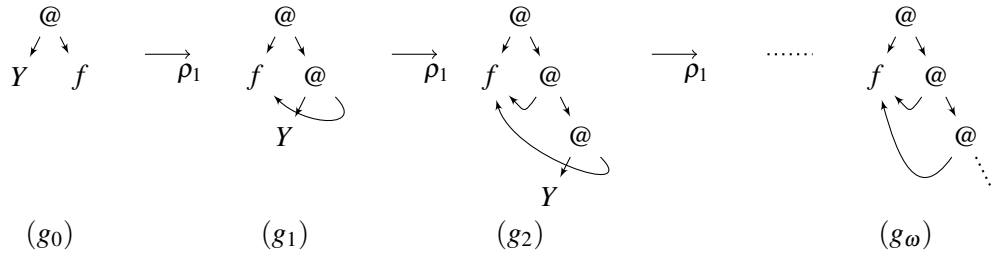
\begin{figure}
  \centering%
  \subfloat[Term graph rules that unravel to $Y\, x \rightarrow x\, (Y\,
  x)$.]{%
    \label{fig:fixedPointCombA}%
    \begin{tikzpicture}%
      \node[node name=180:\underline{l}] (l) {$@$}%
      child {%
        node (Y) {$Y$}%
      } child {%
        node (f) {$x$}%
      };%
      \node[node distance=1.3cm,right=of l,node name=180:\underline{r}] (r) {$@$}%
      child[missing]%
      child {%
        node (a) {$@$}%
        child {%
          node {$Y$}%
        } child[missing]
      };%
      \draw%
      (r) edge[->,out=-135,in=35] (f)
      (a) edge[->,out=-45,in=-35] (f);%
      \node at ($(l)!.5!(r) + (0,-2)$) {$(\rho_1)$};
    \end{tikzpicture}
    \hspace{1cm}
    \begin{tikzpicture}
      \node[node name=180:\underline{l}] (l) {$@$}%
      child {%
        node (Y) {$Y$}%
      } child {%
        node (f) {$x$}%
      };%
      \node[node distance=1.3cm,right=of l,node name=180:\underline{r}] (r) {$@$};%
      \draw%
      (r) edge[->,out=-45,in=-15] (l);%
      \node at ($(l)!.5!(r) + (0,-2)$) {$(\rho_2)$};
      \draw%
      (r) edge[->,out=-135,in=35] (f);%
    \end{tikzpicture}%
  }%
  \quad
  \subfloat[A single $\rho_2$-step.]{%
    \label{fig:fixedPointCombB}%
    \begin{tikzpicture}
      \node (g1) {$@$}%
      child {%
        node {$Y$}%
      } child {%
        node {$f$}
        };%

      \node at ($(g1) + (0,-2)$) {$(g_0)$};      
      \node [node distance=2.5cm,right=of g1] (g2) {$@$}%
      child {%
        node {$f$}%
      } child [missing];%
      \draw (g2) edge[->, out=-45,in=0,loop] (g2);%

      \node at ($(g2) + (0,-2)$) {$(h_0)$};
      \node (s1) at ($(g1)!.5!(g2)-(0,.5)$) {};

      \draw[single step] ($(s1)-(.3,0)$) -- ($(s1)+(.3,0)$)
      node[pos=1,below] {{\small$\rho_2$}};
    \end{tikzpicture}
  }%

    \subfloat[A strongly $\mrs$-convergent term graph reduction over
    $\rho_1$.]{%
    \label{fig:fixedPointCombC}%
      \begin{tikzpicture}  
      \node (g1) {$@$}%
      child {%
        node (a) {$Y$}%
      } child {%
        node {$f$}
      };%

      \node at ($(g1) + (0,-2.9)$) {$(g_0)$};
        
      \node [node distance=2.5cm,right=of g1] (g2) {$@$}%
      child {%
        node (f) {$f$}%
      } child {%
        node (c) {$@$}%
        child {%
          node (a) {$Y$}%
        } child [missing]%
      };%
      \draw (c) edge[->, min distance=6mm,out=-45,in=-45] (f);%

      \node () at ($(g2) + (0,-2.9)$) {$(g_1)$};
        
      \node [node distance=2.5cm,right=of g2] (g3) {$@$}%
      child {%
        node (f) {$f$}%
      } child {%
        node (c) {$@$}%
        child [missing]%
        child {%
          node (c2) {$@$}%
          child {%
            node (a) {$Y$}%
          } child [missing]%
        }%
      };%
      \draw (c) edge[->, min distance=3mm,out=-125,in=-45] (f);%
      \draw (c2) edge[->, min distance=8mm,out=-45,in=-75] (f);%

      \node () at ($(g3) + (0,-2.9)$) {$(g_2)$};
        
      \node [node distance=4.5cm,right=of g3] (go) {$@$}%
      child {%
        node (f) {$f$}%
      } child {%
        node (c) {$@$}%
        child [missing]%
        child {%
          node (c2) {$@$}%
          child [missing]%
          child[etc] {%
            node {} %
          }%
        }
      };%
      \draw (c) edge[->, min distance=3mm,out=-125,in=-45] (f);%
      \draw (c2) edge[->, min distance=8mm,out=-125,in=-75] (f);%

      \node () at ($(go) + (0,-2.9)$) {$(g_\omega)$};

    \node (s1) at ($(g1)!.5!(g2)-(0,.5)$) {};
    \node (s2) at ($(g2)!.55!(g3)-(0,.5)$) {};
    \node (s3) at ($(g3)!.4!(go)-(0,.5)$) {};
    \node (s4) at ($(g3)!.7!(go)-(0,.5)$) {};

    \draw[single step] ($(s1)-(.3,0)$) -- ($(s1)+(.3,0)$)
    node[pos=1,below] {{\small$\rho_1$}};
    \draw[single step] ($(s2)-(.3,0)$) -- ($(s2)+(.3,0)$)
    node[pos=1,below] {{\small$\rho_1$}};
    \draw[single step] ($(s3)-(.3,0)$) -- ($(s3)+(.3,0)$)
    node[pos=1,below] {{\small$\rho_1$}};
    \draw[dotted,thick,-] ($(s4)-(.3,0)$) -- ($(s4)+(.3,0)$);
        
    \end{tikzpicture}
  }%
  \caption{Implementation of the fixed point combinator as a term
    graph rewrite rule.}
  \label{fig:fixedPointComb}
\end{figure}

\begin{example}
  \label{ex:fixedPointCombRules}
  Figure~\ref{fig:fixedPointCombA} shows two term graph rules which
  both unravel to the term rule $\rho\fcolon @(Y, x) \to @(x,@(Y,x))$
  that defines the fixed point combinator $Y$. Note that sharing of
  nodes is used both to refer to variables in the left-hand side from
  the right-hand side and in order to simulate duplication.
\end{example}

Without going into all details of the construction, we describe the
application of a rewrite rule $\rho$ with root nodes $l$ and $r$ to a
term graph $g$ in four steps: at first a suitable sub-term graph of
$g$ rooted in some node $n$ of $g$ is \emph{matched} against the
left-hand side of $\rho$. This matching amounts to finding a
$\calV$-homomorphism $\phi$ from the left-hand side $\lhs\rho$ to
$\subgraph{g}{n}$, the \emph{redex}. The $\calV$-homomorphism $\phi$
allows us to instantiate variables in the rule with sub-term graphs of
the redex. In the second step, nodes and edges in $\rho$ that are not
in $\lhs\rho$ are copied into $g$, such that each edge pointing to a
node $m$ in $\lhs\rho$ is redirected to $\phi(m)$. In the next step,
all edges pointing to the root $n$ of the redex are redirected to the
root $n'$ of the \emph{contractum}, which is either $r$ or $\phi(r)$,
depending on whether $r$ has been copied into $g$ or not (because it
is reachable from $l$ in $\rho$). Finally, all nodes not reachable
from the root of (the now modified version of) $g$ are removed.  With
$h$ the result of the above construction, we obtain a
\emph{pre-reduction step} $\psi\fcolon g \preto[n] h$ from $g$ to $h$.

The definition of term graph rewriting in the form of pre-reduction
steps is very operational. While this style is beneficial for
implementing a rewriting system, it is problematic for reasoning on
term graphs modulo isomorphism, which is necessary for introducing
notions of convergence. However, one can easily see that the
construction of the result term graph of a pre-reduction step is
invariant under isomorphism, which justifies the following definition
of reduction steps:
\begin{definition}
  Let $\calR = (\Sigma,R)$ be GRS, $\rho \in R$ and $g,h \in
  \ictgraphs$ with $n \in N^g$ and $m\in N^h$. A tuple $\phi =
  (g,n,h)$ is called a \emph{reduction step}, written $\phi\fcolon g
  \to[n] h$, if there is a pre-reduction step $\phi'\fcolon g'
  \preto[n'] h'$ with $\canon{g'} = g$, $\canon{h'} = h$, and $n =
  \nodePos{g'}{n'}$. We also write $\phi\fcolon g \to[\calR] h$ to
  indicate $\calR$.
\end{definition}
In other words, a reduction step is a canonicalised pre-reduction
step. Figure~\ref{fig:fixedPointCombB} and
Figure~\ref{fig:fixedPointCombC} illustrate some (pre-)reduction steps
induced by the rules $\rho_1$ respectively $\rho_2$ shown in
Figure~\ref{fig:fixedPointCombA}.


\subsection{Weak Convergence}
\label{sec:weak-convergence}

In analogy to infinitary term rewriting, we employ the partial order
$\lebots$ and the metric $\dds$ for the purpose of defining
convergence of transfinite term graph reductions.
\begin{definition}
  Let $\calR = (\Sigma,R)$ be a GRS.
  \begin{enumerate}[(i)]
  \item Let $S = (g_\iota \to_\calR g_{\iota+1})_{\iota < \alpha}$ be
    a reduction in $\calR$. $S$ is \emph{weakly $\mrs$-continuous} in
    $\calR$ if $\lim_{\iota\limto\lambda} g_\iota = g_\lambda$ for
    each limit ordinal $\lambda < \alpha$. $S$ \emph{weakly
      $\mrs$-converges} to $g \in \ictgraphs$ in $\calR$, written
    $S\fcolon g_0 \wmato[\calR] g$, if it is weakly $\mrs$-continuous
    and $\lim_{\iota\limto\wsuc\alpha} g_\iota = g$.
  \item Let $\calR_\bot$ be the GRS $(\Sigma_\bot, R)$ over the
    extended signature $\Sigma_\bot$ and $S = (g_\iota \to[\calR_\bot]
    g_{\iota+1})_{\iota < \alpha}$ a reduction in $\calR_\bot$. $S$ is
    \emph{weakly $\prs$-continuous} in $\calR$ if
    $\liminf_{\iota<\lambda} g_i = g_\lambda$ for each limit ordinal
    $\lambda < \alpha$. $S$ \emph{weakly $\prs$-converges} to
    $g\in\ipctgraphs$ in $\calR$, written $S\fcolon g_0 \wpato[\calR]
    g$, if it is weakly $\prs$-continuous and
    $\liminf_{\iota<\wsuc\alpha} g_i = g$.
  \end{enumerate}
\end{definition}

\begin{example}
  \label{ex:fixedPointCombWeak}
  Figure \ref{fig:fixedPointCombC} illustrates an infinite reduction
  derived from the rule $\rho_1$ in Figure~\ref{fig:fixedPointCombA}.
  Since $\truncs{g_i}{(i+1)} \isom \truncs{g_\omega}{(i+1)}$ for all
  $i < \omega$, we have that $\lim_{i \limto \omega} g_i = g_\omega$,
  which means that the reduction weakly $\mrs$-converges to the term
  graph $g_\omega$. Moreover, since each node in $g_\omega$ eventually
  appears in a term graph in $(g_i)_{i<\omega}$ and remains stable
  afterwards, we have $\liminf_{i\limto\omega}g_\iota =
  g_\omega$. Consequently, the reduction also weakly $\prs$-converges
  to $g_\omega$.
\end{example}

Recall that weak $\prs$-convergence for TRSs is a conservative
extension of weak $\mrs$-convergence (cf.\
Theorem~\ref{thr:strongExt}). The key property that makes this
possible is that for each sequence $(t_\iota)_{\iota<\alpha}$ in
$\iterms$, we have that $\lim_{\iota\limto\alpha} t_\iota =
\liminf_{\iota\limto\alpha} t_\iota$ whenever
$(t_\iota)_{\iota<\alpha}$ converges, or $\liminf_{\iota\limto\alpha}
t_\iota$ is a total term. Sadly, this is not the case for the metric
space and the partial order on term graphs: the sequence of term
graphs depicted in Figure~\ref{fig:convWeird} has a total term graph
as its limit inferior, viz.\ $g_\omega$, although it does not converge
in the metric space. In fact, since the sequence in
Figure~\ref{fig:convWeird} alternates between two distinct term
graphs, it does not converge in any Hausdorff space, i.e.\ in
particular, it does not converge in any metric space.

This example shows that we cannot hope to generalise the compatibility
property that we have for terms: even if a sequence of total term
graphs has a total term graph as its limit inferior, it might not
converge. However, the converse direction of the correspondence does
hold true:
\begin{theorem}
  \label{thr:limLiminf}
  If $(g_\iota)_{\iota<\alpha}$ converges, then
  $\lim_{\iota\limto\alpha} g_\iota = \liminf_{\iota\limto\alpha}
  g_\iota$.
\end{theorem}

From this property, we obtain the following relation between weak
$\mrs$- and $\prs$-convergence:
\begin{theorem}
  Let $S$ be a reduction in a GRS $\calR$.
  $\text{If}\quad S\fcolon g \wmato[\calR] h \qquad \text{then} \qquad
  S\fcolon g \wpato[\calR] h.$
\end{theorem}
As indicated above, weak $\mrs$-convergence is not the total fragment
of weak $\prs$-convergence as it is the case for TRSs, i.e.\ the
converse of the above implication does not hold in general:
\begin{example}
  \label{ex:rulesWeird}
  There is a GRS that yields the reduction shown in
  Figure~\ref{fig:convWeird}, which weakly $\prs$-converges to
  $g_\omega$ but is not weakly $\mrs$-convergent. This reduction can
  be produced by alternately applying the rules $\rho_1,\rho_2$, where
  the left hand side of both rules and the right-hand side of $\rho_1$
  is $g_0$, and the right-hand side of $\rho_2$ is $g_1$.
\end{example}

\subsection{Strong Convergence}
\label{sec:strong-convergence}

The idea of strong convergence is to conservatively approximate the
convergence behaviour somewhat independently from the actual rewrite
rules that are applied. Strong $\mrs$-convergence in TRSs requires
that the depth of the redexes tends to infinity thereby assuming that
anything at the depth of the redex or below is potentially affected by
a reduction step. Strong $\prs$-convergence, on the other hand, uses a
better approximation that only assumes that the redex is affected by a
reduction step -- not however other subterms at the same depth. To
this end strong $\prs$-convergence uses a notion of reduction contexts
-- essentially the term minus the redex -- for the formation of
limits. The following definition provides the construction for the
notion of reduction contexts that we shall use for term graph
rewriting:
\begin{definition}
  \label{def:truncl}
  Let $g \in \iptgraphs$ and $n \in N^g$. The \emph{local truncation}
  of $g$ at $n$, denoted $\truncl{g}{n}$, is obtained from $g$ by
  labelling $n$ with $\bot$ and removing all outgoing edges from $n$
  as well as all nodes that thus become unreachable from the
  root.
\end{definition}

\begin{proposition}
  \label{prop:stepContext}
  Given a reduction step $g \to[n] h$, we have
  $\truncl{g}{n} \lebots g, h$.
\end{proposition}
This means that the local truncation at the root of the redex is
preserved by reduction steps and is therefore an adequate notion of
reduction context for strong
$\prs$-convergence~\cite{bahr10rta}. Using this construction we can
define strong $\prs$-convergence on term graphs analogously to strong
$\prs$-convergence on terms. For strong $\mrs$-convergence, we simply
take the same notion of depth that we already used for the definition
of the simple truncation $\truncs{g}{d}$ and thus the simple metric
$\dds$.
\begin{definition}
  Let $\calR = (\Sigma,R)$ be a GRS.
  \begin{enumerate}[(i)]
  \item The \emph{reduction context} $c$ of a graph reduction step
    $\phi\fcolon g \to[n] h$ is the term graph
    $\canon{\truncl{g}{n}}$. We write $\phi\fcolon g \to[c] h$ to
    indicate the reduction context of a graph reduction step.
  \item Let $S = (g_\iota \to[n_\iota] g_{\iota+1})_{\iota<\alpha}$ be
    a reduction in $\calR$. $S$ is \emph{strongly $\mrs$-continuous}
    in $\calR$ if $\lim_{\iota \limto \lambda} g_\iota = g_\lambda$
    and $(\depth{g_\iota}{n_\iota})_{\iota<\lambda}$ tends to infinity
    for each limit ordinal $\lambda < \alpha$. $S$ \emph{strongly
      $\mrs$-converges} to $g$ in $\calR$, denoted $S\fcolon g_0
    \mato[\calR] g$, if it is strongly $\mrs$-continuous and either
    $S$ is closed with $g = g_\alpha$ or $S$ is open with $g =
    \lim_{\iota \limto \alpha} g_\iota$ and
    $(\depth{g_\iota}{n_\iota})_{\iota<\alpha}$ tending to infinity.
  \item Let $S = (g_\iota \to[c_\iota] g_{\iota+1})_{\iota<\alpha}$ be
    a reduction in $\calR_\bot=(\Sigma_\bot,R)$. $S$ is \emph{strongly
      $\prs$-continuous} in $\calR$ if $\liminf_{\iota \limto \lambda}
    c_\iota = g_\lambda$ for each limit ordinal $\lambda <
    \alpha$. $S$ \emph{strongly $\prs$-converges} to $g$ in $\calR$,
    denoted $S\fcolon g_0 \pato[\calR] g$, if it is strongly
    $\prs$-continuous and either $S$ is closed with $g = g_\alpha$ or
    $S$ is open with $g = \liminf_{\iota \limto \alpha} c_\iota$.
  \end{enumerate}
\end{definition}

\begin{example}
  As explained in Example~\ref{ex:fixedPointCombWeak}, the reduction
  in Figure~\ref{fig:fixedPointCombC} both weakly $\mrs$- and
  $\prs$-converges to $g_\omega$. Because contraction takes place at
  increasingly large depth, the reduction also strongly
  $\mrs$-converges to $g_\omega$. Moreover, since each node in
  $g_\omega$ eventually appears also in the sequence of reduction
  contexts $(c_i)_{i<\omega}$ of the reduction and remains stable
  afterwards, we have that $\liminf_{i\limto\omega}c_i =
  g_\omega$. Consequently, the reduction also strongly
  $\prs$-converges to $g_\omega$.
\end{example}

Remarkably, one of the advantages of the strong variant of convergence
is that we regain the correspondence between $\mrs$- and
$\prs$-convergence that we know from infinitary term rewriting:
\begin{theorem}[\cite{bahr12rta}]
  \label{thr:graphExt}
  Let $\calR$ be a GRS and $S$ a reduction in $\calR_\bot$. We then have
  that
  \begin{center}
    $S\fcolon g \mato[\calR] h$ \qquad \text{iff} \qquad $S\fcolon g
    \pato[\calR] h$ in $\ictgraphs$.
  \end{center}
\end{theorem}

In particular, the GRS given in Example~\ref{ex:rulesWeird} that
induces the reduction depicted in Figure~\ref{fig:convWeird} does not
provide a counterexample for the ``if'' direction of the above
equivalence -- in contrast to weak convergence. The reduction in
Figure~\ref{fig:convWeird} does not strongly $\mrs$-converge but it
does strongly $\prs$-converge to the term graph $\bot$, which is in
accordance with Theorem~\ref{thr:graphExt} above.

\subsection{Soundness and Completeness}
\label{sec:soundness}

In order to assess the value of the modes of convergence on term
graphs that we introduced in this paper, we need to compare them to
the well-established counterparts on terms. Ideally, we would like to
see a strong connection between converging reductions in a GRS $\calR$
and converging reductions in the TRS $\unrav{\calR}$ in the form of
soundness and completeness properties.  For example, for
$\mrs$-convergence we want to see that $g \wmato[\calR] h$ implies
$\unrav g \wmato[\unrav{\calR}] \unrav h$ -- i.e.\ soundness -- and
vice versa that $\unrav g \wmato[\unrav{\calR}] t$ implies $g
\wmato[\calR] h$ with $\unrav h = t$ -- i.e.\ completeness.

Completeness is already an issue for finitary
rewriting~\cite{kennaway94toplas}: a single term graph redex may
correspond to several term redexes due to sharing. Hence, contracting
a term graph redex may correspond to several term rewriting steps,
which may be performed independently.

In the context of weak convergence, also soundness becomes an
issue. The underlying reason for this issue is similar to the
phenomenon explained above: a single term graph rewrite step may
represent several term rewriting steps, i.e.\ $g \to[\calR] h$ implies
$\unrav g \fto+[\unrav\calR]\unrav h$.\footnote{If the term graph $g$
  is cyclic, the corresponding term reduction may even be infinite.}
When we have a converging term graph reduction $(\phi_\iota\fcolon
g_\iota \to g_{\iota+1})_{\iota<\alpha}$, we know that the underlying
sequence of term graphs $(g_\iota)_{\iota<\wsuc\alpha}$
converges. However, the corresponding term reduction does not
necessarily produce the sequence
$(\unrav{g_\iota})_{\iota<\wsuc\alpha}$ but may intersperse the
sequence $(\unrav{g_\iota})_{\iota<\wsuc\alpha}$ with additional
intermediate terms, which might change the convergence behaviour.

While we cannot prove soundness for weak convergence due to the
abovementioned problems, we can show that the underlying modes of
convergence are sound in the sense that convergence is preserved under
unravelling.
\begin{theorem}
  \label{thr:unravLim}
  \quad
  \begin{enumerate}[(i)]
  \item $\lim_{\iota\limto\alpha} g_\iota = g$ implies
    $\lim_{\iota\limto\alpha} \unrav{g_\iota} = \unrav g$ for every
    sequence $(g_\iota)_{\iota<\alpha}$ in $(\ictgraphs,\dds)$.
  \item $\unrav{\liminf_{\iota\limto\alpha}g_\iota} =
    \liminf_{\iota\limto\alpha}\unrav{g_\iota}$ for every sequence
    $(g_\iota)_{\iota<\alpha}$ in $(\ipctgraphs,\lebots)$.
  \end{enumerate}
\end{theorem}
Note that the above theorem in fact implies soundness of the modes of
convergence on term graphs with the modes of convergence on terms
since both $\dds$ and $\lebots$ specialise to $\dd$ respectively
$\lebot$ if restricted to term trees.

However, we can observe that strong convergence is more well-behaved
than weak convergence. It is possible to prove soundness and
completeness properties for strong $\prs$-convergence:
\begin{theorem}[\cite{bahr12rta}]
  \label{thr:pConvSoundCompl}
  Let $\calR$ be a left-finite GRS.
  \begin{enumerate}[(i)]
  \item If $\calR$ is left-linear and $g \pato[\calR] h$, then
    $\unrav{g} \pato[\unrav{\calR}] \unrav{h}$.
  \item If $\calR$ is orthogonal and $\unrav{g} \pato[\unrav\calR] t$,
    then there are reductions $g \pato[\calR] h$ and $t
    \pato[\unrav\calR] \unrav{h}$.
  \end{enumerate}
\end{theorem}
Note that the above completeness property is not the one that one
would initially expect, namely $\unrav g \pato[\unrav{\calR}] t$
implies $g \pato[\calR] h$ with $\unrav h = t$. But this general
completeness property is known to already fail for finitary term graph
rewriting~\cite{kennaway94toplas}.

The soundness and completeness properties above have an important
practical implication: GRSs that only differ in their sharing, i.e.\
they unravel to the same TRS, will produce the same results, i.e.\ the
same normal forms up to bisimilarity. GRSs with more sharing may,
however, reach a result with fewer steps. This can be observed in
Figure~\ref{fig:fixedPointComb}, which depicts two rules $\rho_1,
\rho_2$ that unravel to the same term rule. Rule $\rho_1$ reaches
$g_\omega$ in $\omega$ steps whereas $\rho_2$ reaches a term graph
$h_0$, which is bisimilar to $g_\omega$, in one step.

The situation for strong $\mrs$-convergence is not the same as for
strong $\prs$-convergence. While we do have soundness under the same
preconditions, i.e.\ $g \mato[\calR] h$ implies $\unrav{g}
\mato[\unrav{\calR}] \unrav{h}$, the completeness property we have
seen in Theorem~\ref{thr:pConvSoundCompl} fails. This behaviour was
already recognised by Kennaway et
al.~\cite{kennaway94toplas}. Nevertheless, we can find a weaker form
of completeness that is restricted to normalising reductions:
\begin{theorem}[\cite{bahr12rta}]
  Given an orthogonal, left-finite GRS $\calR$ that is normalising
  w.r.t.\ strongly $\mrs$-converging reductions, we find for each
  normalising reduction $\unrav{g} \mato[\unrav\calR] t$ a reduction
  $g \mato[\calR] h$ such that $t = \unrav{h}$.
\end{theorem}

\section{Concluding Remarks}
\label{sec:concluding-remarks}

We have devised two independently defined but closely related
infinitary calculi of term graph rewriting. This is not the first
proposal for infinitary term graph rewriting calculi; in previous
work~\cite{bahr12lmcs} we presented a so-called \emph{rigid} approach
based on a metric and a partial order different from the structures
presented here.

There are several arguments why the simple approach presented in this
paper is superior to the rigid approach. First of all it is
simpler. The rigid metric and partial order have been carefully
crafted in order to obtain a correspondence result in the style of
Theorem~\ref{thr:strongExt} for weak convergence on term graphs. This
correspondence result of the rigid approach is not fully matched by
the simple approach that we presented here, but we do regain the full
correspondence by moving to strong convergence.

Secondly, the rigid approach is very restrictive, disallowing many
reductions that are intuitively convergent. For example, in the rigid
approach the reduction depicted in Figure~\ref{fig:fixedPointCombC},
would not $\prs$-converge (weakly or strongly) to the term graph
$g_\omega$ as intuitively expected but instead to the term graph
obtained from $g_\omega$ by replacing $f$ with $\bot$. Moreover, this
sequence would not $\mrs$-converge (weakly or strongly) at all.

Lastly, as a consequence of the restrictive nature of the rigid
approach, the completion constructions of the underlying metric and
partial order do not yield the full set of term graphs -- in contrast
to our findings here in Theorem~\ref{thr:idealCompletion} and
\ref{thr:metricCompletion}.

Unfortunately, we do not have solid soundness or completeness results
for weak convergence apart from the preservation of convergence under
unravelling and the metric/ideal completion construction of the set of
term graphs. However, as we have shown, this shortcoming is again
addressed by moving to strong convergence.

\bibliography{compact.bib}

\end{document}